\documentclass[journal=nalefd,manuscript=article]{achemso}

\usepackage[version=3]{mhchem} 
\usepackage{graphicx}
\usepackage{epstopdf}
\usepackage{dcolumn}
\usepackage{bm}
\usepackage{amsmath}
\usepackage{braket}

\newcommand{\SuppInfo}{supplementary material}

\newcommand{\Eq}[1]{equation~(\ref{#1})}

\newcommand{\Fig}[1]{Fig.~\ref{#1}}
\newcommand{\Figure}[1]{Figure~\ref{#1}}

\author{R. Gaudenzi}
\affiliation{Kavli Institute of Nanoscience, Delft University of Technology, 2600 GA, Delft, The Netherlands}
\author{E. Burzur\'{\i}}
\affiliation{Kavli Institute of Nanoscience, Delft University of Technology, 2600 GA, Delft, The Netherlands}
\email{e.burzurilinares@tudelft.nl}
\author{D. Reta}
\author{I. de P. R. Moreira}
\author{S. T. Bromley}
\affiliation{Departament de Qu\'{\i}mica F\'{\i}sica and Institut de Qu\'{\i}mica Te\'{o}rica i Computacional, Universitat de Barcelona (IQTCUB), E-08028 Barcelona, Spain}
\alsoaffiliation{Institici\'{o} Catalana de Recerca i Estudis Avan\c{c}ats (ICREA), 08010 Barcelona, Spain}
\author{C. Rovira}
\author{J .Veciana}
\affiliation{Institut de Ci\'{e}ncia de Materials de Barcelona (ICMAB-CSIC) and CIBER-BBN, Campus de la UAB, 08193, Bellaterra, Spain}

\author{H. S. J. van der Zant}
\affiliation{Kavli Institute of Nanoscience, Delft University of Technology, 2600 GA, Delft, The Netherlands}

\title{Exchange Coupling Inversion in a High-Spin Organic Triradical Molecule}


\begin{document}
\begin{abstract}
The magnetic properties of a nanoscale system are inextricably linked to its local environment. In ad-atoms on surfaces and inorganic layered structures the exchange interactions result from the relative lattice positions, layer thicknesses and other environmental parameters. Here, we report on a sample-dependent sign inversion of the magnetic exchange coupling between the three unpaired spins of an organic triradical molecule embedded in a three-terminal device. This ferro-to-antiferromagnetic transition is due to structural distortions and results in a high-to-low spin ground state change in a molecule traditionally considered to be a robust high-spin quartet. Moreover, the flexibility of the molecule yields an in-situ electric tunability of the exchange coupling via the gate electrode. These findings open a route to the controlled reversal of the magnetic states in organic molecule-based nanodevices by mechanical means, electrical gating or chemical tailoring.
\end{abstract}

\section{Introduction}

Magnetism at the nanoscale is often determined by the local environment: the sensitivity of the exchange coupling to the spatial arrangement and its dependence on the interactions between sandwiching layers, spin interfaces, ligands, neighbouring atoms or substrate is well-established. In most cases, these interactions lead to an increase or decrease of the strength of the local magnetism \cite{Heinrich2013, Parks2010, Burgess2015, Burzuri2012a, Burzuri2015, Kahn1976}. Of special interest is the situation in which the sign of the exchange interaction reverses, leading to a transition from ferromagnetic to antiferromagnetic coupling (or vice versa). Altering the relative positions of the atoms in Fe dimers\cite{Bryant2013} or varying the thickness of the interlayer in Fe/Cr/Fe structures\cite{Grunberg1986, Fert1995} can, for instance, trigger such an inversion.

Owing to their intrinsic flexibility, single molecules form an interesting system to control this exchange reversal. In addition, as building blocks of molecule-based materials, knowledge on variations of the magnetic properties at the individual-molecule scale reveals effects that might go unseen in bulk. A previous work reports variations of the spin excitation energies in single molecules on metallic surfaces\cite{Burgess2015}, but a ferro-to-antiferromagnetic exchange coupling inversion has never been demonstrated. In view of molecular spintronics applications, such a phenomenon is of particular interest in high-spin all-organic molecules - in which magnetism is not connected to the presence of metal ions\cite{Liu2013,Zhang2013,Mullegger2013,Frisenda2015} - because of their favorable spin lifetimes\cite{Rocha2005}.

Here, via inelastic electron tunneling spectroscopy (IETS), we map the magnetic states of individual neutral and stable organic triradical molecules. The all-organic molecule of our study exhibits, in solution, a strong ferromagnetic exchange interaction between its three unpaired electrons. Through the observation of distinct magnetic spectra in different samples, we infer that the exchange coupling significantly decreases in magnitude and can even turn antiferromagnetic when the molecule is embedded in a solid-state device. We attribute the reduction and the sign reversal to small deformations induced by the local environment of the junction and support this hypothesis with theoretical calculations. The analysis demonstrates that the distortions only modify the exchange but not the robust radical character of the three centers, in agreement with previous studies on monoradicals\cite{Frisenda2015}.

\begin{figure}
\includegraphics[width=.5\columnwidth]{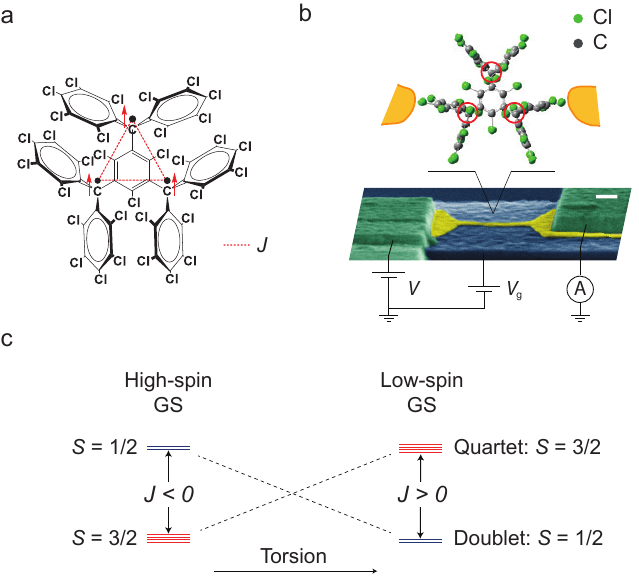}
  \caption{
  \textbf{The molecule, the device and the exchange coupling inversion mechanism.}
  (a) Molecular structure of a triradical organic molecule with a $C_2$ symmetry: the three radical centers, located on the three methyl carbon atoms, are each surrounded by three twisted perchlorinated phenyl rings. The three unpaired electrons are coupled through an exchange interaction, $J$, schematically represented by red dashed lines. (b) Scanning-electron-microscope micrograph (100-nm scale bar, false color) of the gold nanowire on a AuPd/Al$_2$O$_3$ gate. The molecular junction is created with the molecule bridging the nanogap formed during electromigration. (c) Schematics of the exchange coupling sign flipping mechanism: a torsion applied to the molecule increases the exchange coupling from negative (ferro-) to positive values (anti-ferromagnetic) through zero inducing a change from the $S = 3/2$ high-spin ground state (GS) to a $S = 1/2$ low-spin ground state.}
\label{Fig_1}
\end{figure}

We use a 2,4,6-hexakis(pentachlorophenyl) mesityltriyl radical molecule\cite{Veciana1993,Sedo1998} sandwiched between two electromigrated\cite{Burzuri2015} gold leads to construct our molecular junction, as schematically depicted in \Fig{Fig_1}b (see \SuppInfo~for additional information on the molecule and the junction preparation). The molecule, shown in \Fig{Fig_1}a, is a neutral triradical with three unpaired electrons on the three methyl carbon atoms.
Each one of these atoms, with three chlorinated phenyl rings surrounding it in a propeller-like conformation, forms one of the three elementary radical subunits. The central mesitylene ring is common to the three subunits and is used to magnetically interconnect them. Two of the propellers have the same sense of rotation while the third rotates with an opposite sense, conferring the  molecule a $C_2$ symmetry. Owing to this architecture and particular orbital topology \cite{Longuet-Higgins1950, Ovchinnikov1978,Borden1977}, the three radical electrons lie in three distinct non-disjoint, quasi-degenerate, non-bonding singly-occupied molecular orbitals (SOMOs) exhibiting a strong exchange interaction. Previous experiments\cite{Veciana1993} have demonstrated a robust high-spin quartet ($S_Q = 3/2$) ground state with a low-spin doublet ($S_D = 1/2$) excited state well-separated in energy ($|E(\ket{Q})-E(\ket{D})|\gg k_{\text{B}} T$). This characterization is, however, performed on crystals, where the molecules adopt the thermodynamically most stable conformation with a $C_2$ symmetry.

\section{Results and discussion}

\begin{figure}
\includegraphics[width=1\columnwidth]{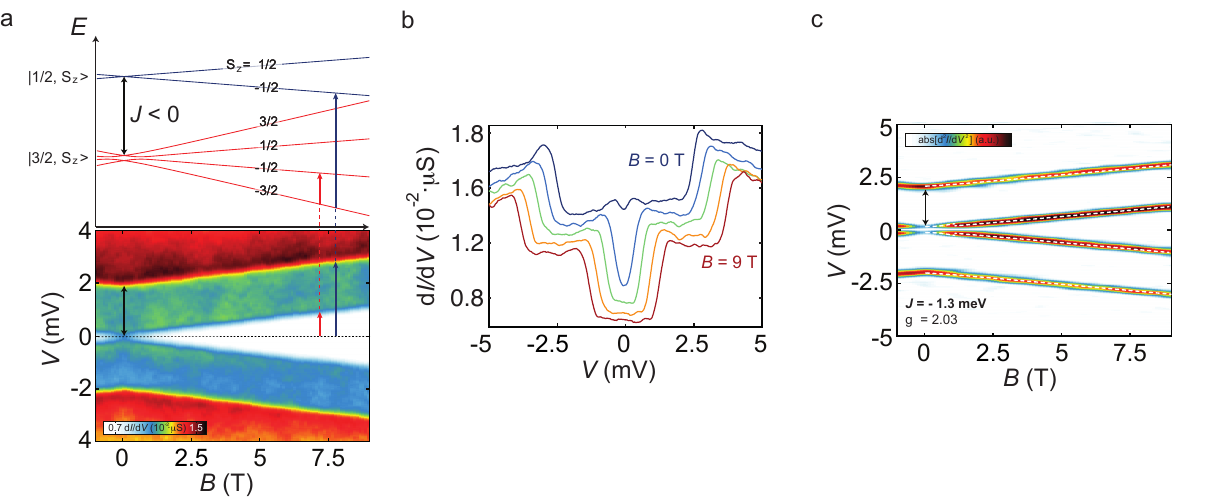}
  \caption{
  \textbf{Ferromagnetic exchange coupling $J$.}
(color online)
(a) $\text{d}I/\text{d} V$ map (below) measured on sample A as a function of bias voltage and magnetic field. The energy splitting between the low- and the high-energy step is constant in magnetic field and marked by a black double arrow.
Above, the energy level scheme of a spin-$3/2$ system with a ferromagnetic exchange coupling $(J < 0)$ and a small anisotropy parameter ($D = 0.06$ meV) as a function of magnetic field. Red and blue arrows indicate the allowed first-order spin-flip processes with $\Delta S_z = \pm 1$ to which the observed steps in $\text{d}I/\text{d} V$ are ascribed.
(b) $\text{d}I/\text{d} V$ spectra extracted from the map in (a) with a spacing of $\Delta B$ = 1.8 T starting from 0 T (offset for clarity). The low-energy excitation exhibits a zero-field splitting of about 0.1 meV.
(c) Absolute value $|\text{d}^2I/\text{d}V^2|$ of the map in (a). The dashed lines superimposed to the experimental data are the fit to the Hamiltonian of equation(1) with $J = -1.3 $ meV $(E(\ket{Q})-E(\ket{D}) = 3/2 J$), $g = 2.03$ and $D = 0.06$ meV. The zero-field spitting is clearly visible. All measurements are taken at $T \approx 70$ mK.}
\label{Fig_2}
\end{figure}

Electron transport spectroscopy on sample A is presented in \Fig{Fig_2} as a function of bias voltage and magnetic field $B$. Four well-defined conductance steps, symmetrically placed at positive and negative bias, are visible in the $\text{d}I/\text{d} V$ color map of \Fig{Fig_2}a and extracted spectra in \Fig{Fig_2}b. The low- and high-energy steps are located around $\pm 0.1$ meV and $\pm 2$ meV at $B = 0$ T and shift linearly and parallel to each other as the magnetic field is increased. The small low-bias step visible at $B = 0$ T signals the presence of a small zero-field splitting ($\approx$ 0.15 meV).

Each finite bias step is associated with the opening of an inelastic electron current channel via an excited state of the molecule. When spin excited states are involved, the steps' position in energy as a function of magnetic field provides a means to read out the molecule's energy spectrum. In the present case, the spectrum is composed of the eigenstates $\ket{\mathbf{S}, S_z}$ of the spin Hamiltonian including exchange, Zeeman and anisotropy terms:
\begin{equation}\label{eq:1}
\hat{\mathcal{H}} = \frac{J}{2}\left(\mathbf{S}^2 - \sum_{i = 1}^3 \mathbf{S}_i^2 \right) + g\mu_\text{B} \mathbf{B}\cdot \mathbf{S} - D S_z^2,
\end{equation}
where $\mathbf{S}_i$ denotes the 1/2-spin vectors of the three radical electrons, $\mathbf{S} \equiv \sum^3_{i=1} \mathbf{S}_i$ and $S_z$ the total spin and spin projection operators, respectively, and $D$ the uniaxial anisotropy parameter.
The second-order spin excitations induced by the tunneling electrons obey the selection rules $\Delta S_z = 0, \pm 1$. Within this framework, we can respectively assign the low- and high-bias steps seen in \Fig{Fig_2}a (bottom) to the two spin transitions $\ket{3/2, -3/2} \:\rightarrow \ket{3/2, -1/2}$ and $\ket{3/2, -3/2} \:\rightarrow \ket{1/2, -1/2}$ between states of the molecular spectrum in \Fig{Fig_2}a (top). The former transition takes place within the spin-3/2 ground state multiplet and approaches therefore zero energy at vanishing magnetic fields; the latter transition, on the other hand, involves an excited state belonging to the higher spin-1/2 multiplet and converges to a finite energy at $B = 0$ T.

In \Fig{Fig_2}c the absolute value $|\text{d}^2I/\text{d}V^2|$ of the map in \Fig{Fig_2}a is shown. The dashed line is a fit to the spin Hamiltonian in \Eq{eq:1} with an exchange coupling $J = - 1.3$ meV and gyromagnetic ratio $g = 2.03$. This negative (ferromagnetic) $J$ favors the high-spin quartet $\ket{3/2, S_z = \pm 3/2, \pm 1/2}$ over the low-spin doublet $\ket{1/2, S_z = \pm 1/2}$. The allowed transitions involve therefore $\Delta S_z = \pm 1$, giving rise to the two inelastic steps in the spectrum that increase linearly with magnetic field.
A weak non-linearity of the low- and high-bias step evolution for fields $B < 0.5$ T is visible. This can be accounted for by setting a non-zero small anisotropy parameter to the spin Hamiltonian of \Eq{eq:1}.

On this sample, gate-dependent measurements have also been performed and are shown in figure S1 of the \SuppInfo. Throughout the entire accessible gate range no sign of resonant transport is seen. This suggests a large SOMO-SUMO (singly-unoccupied molecular orbital) gap and supports charge neutrality. While no resonant transport is visible, we observe that an increase in gate voltage results in a sizeable increase of the exchange coupling. This electric field-induced modulation of $J$ amounts to a $9\%$ of its total value.

We further note that two other measured samples show a similar set of transitions (see \SuppInfo). The extracted exchange coupling constants are in those cases $J = - 2.2$ meV and $J = -2.3$~meV.

\begin{figure}
\includegraphics[width=1\columnwidth]{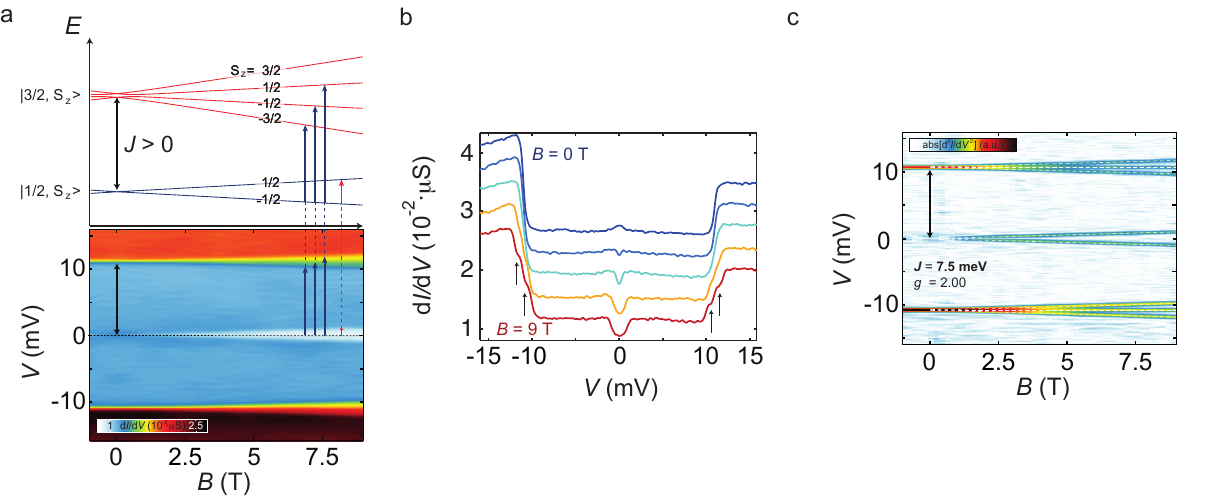}
  \caption{
  \textbf{Anti-ferromagnetic exchange coupling $J$.}
  (color online)
(a) $\text{d}I/\text{d} V$ map measured on sample B as a function of bias voltage and magnetic field. A zero-bias peak and a step are visible at zero magnetic field. At high fields the high energy step is split into three smaller steps, with negative, zero and positive slopes. The upper panel shows the energy level scheme of a spin-$3/2$ system with an antiferromagnetic exchange coupling $(J > 0)$ and a small anisotropy parameter ($D = 0.06$ meV). The colored arrows indicate the allowed processes with $\Delta S_z = 0, \pm 1$ associated with the observed spectrum.
(b) $\text{d}I/\text{d} V$ spectra extracted from the map in (a) with a spacing of $\Delta B$ = 1.8 T between 0 T and 9 T (offset for clarity). As the magnetic field increases, the zero-bias peak splits and the two additional steps appear in correspondence of the single step (black arrows).
(c) Absolute value $|\text{d}^2I/\text{d}V^2|$ of the map in (a) with the fit to equation (1) superimposed (dashed lines). The splitting of the zero-bias peak and higher-energy excitation in three smaller staircase-like excitations is clearly visible. The extracted fitting parameters are $J = 7.5 $ meV, $g = 2.0$ and $D = 0.06$ meV. All measurements are taken at $T \approx 70$ mK.}
\label{Fig_3}
\end{figure}

A second group of samples showed a markedly different set of spin transitions and magnetic field evolution. The characteristics are summed up for sample B in \Fig{Fig_3}. The color map in \Fig{Fig_3}a displays the $\text{d}I/\text{d} V$ as a function of bias voltage and magnetic field. A small zero-bias peak and two symmetric conductance steps at about $\pm 11$ meV are visible at $B = 0$ T. The zero-bias peak evolves into two steps for increasing magnetic fields and the single high-bias step splits in three smaller steps. The three-fold excitation can be more readily seen in \Fig{Fig_3}b (black arrows) and \Fig{Fig_3}c. This last feature is not compatible with the spin-3/2 ground state observed in sample A and indicates a ground state spin $S < 3/2$ together with a excited multiplet with $S+1 \le 3/2 $ (either $S = 1/2$ or $0$).

The presence of the zero-bias peak and its doublet-like magnetic field evolution indicate a $S = 1/2$ ground state. The energy spectrum corresponding to this case is displayed in  \Fig{Fig_3}a (top). The low-bias step is ascribed to the transition $\ket{1/2, -1/2} \:\rightarrow \ket{1/2, 1/2}$ within the spin-1/2 ground state multiplet; the high-bias ones are associated with the three allowed transitions to the spin-3/2 multiplet $\ket{1/2, -1/2} \:\rightarrow \ket{3/2, S_z}$, with $S_z = \{-3/2, -1/2, 1/2\}$, where the selection rule $\Delta S_z = 0$ also applies. Selected spectra extracted from \Fig{Fig_3}a at different fields are displayed in \Fig{Fig_3}b. The zero-bias peak is clearly visible in the trace at zero magnetic field. The peak evolves into a dip at $B \approx 1.8$ T and opens up into two inelastic steps at higher fields.
The observed weak zero-bias peak is consistent with the presence of Kondo correlations between one of the SOMO unpaired electrons of the spin-1/2 ground state and the electrons in the leads.

Taking the absolute value of the derivative of the map in \Fig{Fig_3}a, we obtain the $|\text{d}^2I/\text{d}V^2|$ map of \Fig{Fig_3}c. The splitting with magnetic field in two and three distinct steps of the low- and high-energy excitations respectively is clearly visible. Superimposed we show the result of a fit to \Eq{eq:1}, from which a magnetic exchange coupling $J = 7.5$ meV and $g = 2.0$ are extracted.

\begin{figure}
\includegraphics[width=.4\columnwidth]{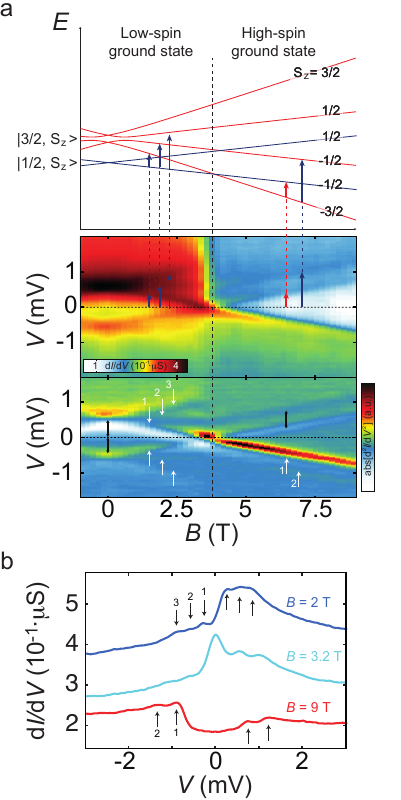}
  \caption{
  \textbf{Magnetic field control over the ground state.} (color online)
(a) $\text{d}I/\text{d} V$ map (center) measured on sample C as a function of $V$ and $B$. At low magnetic fields three conductance excitations (indicated by numbers) split in magnetic field. The positive and negative-bias ones numbered 1 cross at $B \approx 3.2$ T (black dashed line). For $B > 3.2$ T, the three lines merge into two steps. The constant energy splitting between the low- and high-energy step is marked by black arrows in the $\text{d}^2I/\text{d}V^2$ map (below).
Above, the energy diagram of a spin-$3/2$ system with an antiferromagnetic exchange coupling $(J > 0)$ and a small anisotropy parameter as a function of magnetic field. A positive but small $J$ favors the low-spin doublet at low magnetic fields and the high-spin quartet at high magnetic fields. Red and blue arrows indicate the spin-flip processes with $\Delta S_z = 0, \pm 1$.
(b) $\text{d}I/\text{d} V$ spectra extracted from the map in (a) at $B = 2, 3.2$ and 9 T (offset for clarity). The three steps (arrows 1, 2 and 3) present for values $B < 3.2$~T evolve into two steps (arrows 1 and 2) for $B > 3.2$ T.}
\label{Fig_4}
\end{figure}

We have observed a similar magnetic field dependence in two other samples with magnetic exchanges of $J = + 3.0$ meV and $J = + 0.4$ meV (sample C, shown in \Fig{Fig_4}). Thus, in these samples, in contrast to sample A, the exchange coupling $J$ is antiferromagnetic (positive) and stabilizes the low-spin doublet over the high-spin quartet.

Measurements in gate performed on sample B are shown in figure S1 of the \SuppInfo. No sign of charging and resonant transport are present. Analogously to sample A, this indicates a large gap and supports charge neutrality. The electric field-induced modulation of $J$ amounts here to a $2\%$ of its total value.

\Figure{Fig_4} shows the results of a measurement on sample C displaying an intermediate scenario between the ones observed in sample A and B. The $\text{d}I/\text{d} V$ and corresponding $|\text{d}^2I/\text{d}V^2|$ color maps of \Fig{Fig_4}(a) exhibit two conductance steps centered around $\approx \pm 0.55$ meV at $B = 0$ T. For magnetic fields below 3.2 T the steps at positive and negative bias split each into three excitations with positive, zero and negative slopes (numbered 1, 2 and 3 in the $|\text{d}^2I/\text{d}V^2|$ map and the $\text{d}I/\text{d} V$ linecut of \Fig{Fig_4}(b)). The two excitations 1 in the positive and negative bias region intersect at $V = 0$ V and $B \approx 3.2$ T (dashed line in \Fig{Fig_4}(a)). For $B > 3.2 $ T only two of the three excitations survive. The zero-bias one emerges as a prosecution of excitation 1 with the same slope, while the finite-bias one continues from 2 with a different slope. The energy difference between the two steps is constant with magnetic field and amounts to $ \pm 0.55$ meV. It is important to notice that this value equals the energy of the excitation at zero magnetic field (black arrows in the map of \Fig{Fig_4}(a)(bottom)). Following the energy level scheme of \Fig{Fig_4}(a), the three excitations in the low-field region ($B < 3.2 $ T) of the plot are associated with the transitions $\ket{1/2, -1/2} \:\rightarrow \ket{3/2, S_z}$, with $S_z = \{-3/2, -1/2, 1/2\}$. The low- and high-bias ones in the high-field side ($B > 3.2 $ T) are ascribed to the transitions $\ket{3/2, -3/2} \:\rightarrow \ket{3/2, -1/2}$ and $\ket{3/2, -3/2} \:\rightarrow \ket{1/2, -1/2}$ respectively. A change from low- to high-spin ground state thus occurs at the crossing point of the two regions ($B = 3.2 $ T).

The set of transitions featured in this sample C can be explained with a positive, but small ($J \sim g\mu_\text{B} B$) exchange coupling $J$. This antiferromagnetic coupling favors the low spin state, but only up to a magnetic field equal to $\approx \frac{2}{3} J/g\mu_\text{B} $. For higher fields a spin-1/2 to spin-3/2 ground state flip occurs, with a consequent change in the excitation spectrum. Owing to the small $J$, the magnetic field provides thus a means to effectively control the magnetic ground state. Importantly, the spectra of samples A and B are therefore shown to be connected exclusively via a magnetic field change, with no need for oxidation/reduction of the molecule.

\begin{figure}
\includegraphics[width=1\columnwidth]{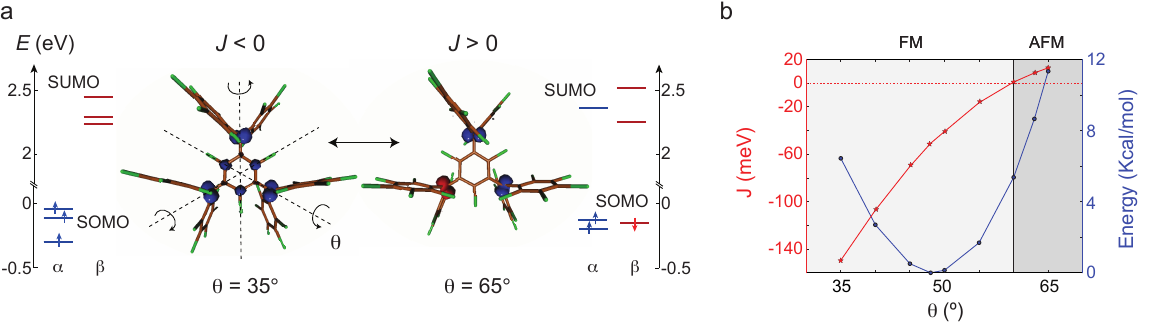}
  \caption{
  \textbf{Modelling orbitals energies and magnetic exchange coupling.}
  (color online) (a) Energy level diagram, molecular structure and spin density isosurfaces for two distinct torsion angles $\theta$. A torsion applied to the three peripheral groups with respect to the central ring promotes the flip of one the spins (rightmost diagram) and a further concentration of the spin density onto the orbitals of methyl carbon atoms. (b) Energy and exchange coupling vs. angle $\theta$ plot. At the potential energy minimum ($\theta = 47^\circ$) the ferromagnetic exchange energy exceeds room-T. Increasing (decreasing) the torsion angle results into an decrease (increase) of $|J|$ and ultimately to the reversal of its sign.
}
\label{Fig_5}
\end{figure}

The data show that different samples of the same neutral individual triradical molecule in an electromigrated junction yield values of $J$ spanning from $- 2.3$ meV to $+ 7.5$ meV, in contrast to the robust value $J \leq -40$ meV obtained from the same molecule in crystals\cite{Veciana1993}. Excluding charging effects on the basis of the gate measurements and large calculated SOMO-SUMO (also named SOMO-LUMO-$\beta$) gap, we argue that the local environment of the junction is responsible for the reduction and sign change of $J$. The mechanism we propose relies on a structural distortion induced in the molecule by the electrodes. In particular, we focus on the dihedral angles $\theta$ of the bonds linking the radical centres to the central ring (\Fig{Fig_5}(a) and \SuppInfo) which are known to be of general importance in radical systems of this type\cite{Alcon2015}. We argue that, these three angles which define the relative orientation of the six peripheral rings with respect to the central one, determine the exchange through their dependence on the specific arrangement of the molecule between the electrodes, thus giving rise to the observed sample to sample variation.
To test this hypothesis we perform unrestricted DFT-based calculations based on the broken symmetry approach\cite{RetaManeru2015a, Noodleman1981, Noodleman1986,Noodleman1995,Moreira2006} for different dihedral angles $\theta$. The $J$-value at each step is extracted from the energy difference between the high-spin and the broken symmetry solution approaching the spin adapted doublet state (See \SuppInfo $\:\:$for details on the calculations and explicit definition of the dihedral angle).

Spin-unrestricted molecular orbitals and magnetic exchange values resulting from the calculations are shown in \Fig{Fig_5}. At low angles ($\theta = 35^\circ$) the three spin-up SOMOs present a large energetic separation from the excited LUMO$-\beta$ orbitals. The non-disjoint\cite{Borden1977} and near-degenerate character of these SOMOs is at the origin of the preferential high-spin GS. The spin density associated to this configuration is distributed on each of the three radical centers with a remarkable participation on the central phenyl ring. This through-bond delocalization of the unpaired electrons on the central ring determines the large exchange integral contributing to the stabilization of the high spin quartet. The spin delocalization over the central ring is progressively cut off as a torsion is applied and the dihedral angle increased\cite{Alcon2015}. At high angles ($\theta = 65^\circ$ is taken here as an example) the spin density on the central ring is completely suppressed and the orthogonality of the SOMOs compromised. The resulting large orbital overlap term favors electron pairing over the unpairing due to the exchange integral, yielding the observed positive (AFM) exchange coupling.

Energy cost and exchange coupling as a function of $\theta$ are reported in \Fig{Fig_5}(b). The most thermodynamically stable conformation, obtained for $\theta = 47^\circ$, is associated with a strong FM interaction. The calculated value ($|J| \approx 40$ meV $ > k_\text{b}T$, with $T = 300$ K) is consistent with the high-spin ground state observed in Electron Spin Resonance (ESR) measurements of the molecule in solid state\cite{Veciana1993}. $J$ monotonically increases for higher angles and the crossover to an AFM coupling is observed at $\theta = 60^\circ$ at an energetic cost of only $\sim 5 $ Kcal/mol\footnote{Given the approximate symmetry of the energetic cost as a function of torsion angle, one would expect to observe also configurations with $|J| > 40$ meV. However, similar variations in angle would yield to $J\sim$ 100 meV according to our calculations. These configurations might go unseen in our measurements due to the fact that the bias window is limited to $\pm 30$ meV to ensure the full stability of the gold nanogap.}. This conformation away from equilibrium can be also attained in the special case of fast-frozen solutions (see \SuppInfo). The observed tunability of $J$ with gate voltage can also be explained through this model. Provided unequal distances between the three radical electronic orbitals and the gate electrode, the electrostatic force can generate a net torque on the molecule. The torque can consequently result into a change in $\theta$.

This model likely presents a simplified picture of the high complexity inherent to the molecule-electrodes coupled system where other kinds of distortions away from the thermodynamically most stable conformation may occur. Despite its simplicity, the model clearly demonstrates that non-destructive torsions applied to such polyradicals can lead to an inversion of the sign of the exchange coupling.
Furthermore, the knowledge of this mechanism opens a pathway to the chemical, mechanical and electrical control of the reversal of magnetic states in individual as well as ensembles of these molecules. Such kind of direct control could be attained, for instance, by \textit{in situ}-modifying the angle $\theta$ with the tip of a scanning tunneling microscope or a mechanically-controlled break junction.
From the chemical point of view, the steric hindrance of the substituents can be engineered to stabilize the antiferromagnetic frustrated configuration in a single molecule.

In summary, we demonstrate that an individual high-spin all-organic triradical molecule in a junction exhibits changes in its exchange coupling between FM and AFM, while maintaining chemical integrity and charge neutrality. The change and sign inversion show, as supported by theory, that at the nanoscale the preference for the high-spin vs. low-spin ground state is not only dictated by the peculiar orbital topology arguments but also by molecular distortions imposed by the surrounding environment.  Furthermore, external electric and magnetic fields may effectively control the $J$-value and the magnetic ground state. These results contribute to the understanding of the influence of the environment on the magnetic properties of molecule-based organic materials and open the way to the control of the magnetic ground states of flexible polyradicals for future molecular spintronics applications.

\begin{acknowledgement}

This work was supported by an advanced ERC grant (Mols@Mols). We also acknowledge financial support by the Dutch Organization for Fundamental research (NWO/FOM).  JV, CR, IPRM, DRM, SB thank funds from Networking Research Center on Bioengineering, Biomaterials and Nanomedicine (CIBER-BBN) DGI and MAT2012-30924 (Spain) and from MINECO (BEWELL CTQ2013-40480-R, PRI-PIBIN-2011-1028 and CTQ2012-30751) and Generalitat de Catalunya (2014-SGR-17, 2014-SGR-97, 2014-SGR-97, XRQTC). EB thanks funds from the EU FP7 program through project 618082 ACMOL and the Dutch funding organization NWO (VENI). JV and CR thank funds from the EU project ITN iSwitch (642196). We acknowledge C. Franco and V. Lloveras for checking the sample purity and V. Lloveras (ICMAB) for ESR spectra.
\\

\end{acknowledgement}


\bibliography{Triradicals}

\clearpage

\end{document}